\documentclass[epsf,twocolumn,preprintnumbers]{revtex4}

\usepackage{amsmath,amssymb,amsfonts}
\usepackage{graphics}
\usepackage{dcolumn} 
\usepackage{bm,graphicx,color}
\usepackage{epsfig,comment}
\begin{document}
\newcommand{\lno}{LaNiO$_2$}
\newcommand{\nno}{NdNiO$_2$}
\newcommand{\Sno}{Nd$_{0.8}$Sr$_{0.2}$NiO$_2$}
\newcommand{\lco}{LaCuO$_2$}
\newcommand{\sco}{SrCuO$_2$}
\newcommand{\cco}{CaCuO$_2$}
\newcommand{\scy}{superconductivity}

\title{Role of $4f$ states in infinite-layer NdNiO$_2$}
\author{Mi-Young Choi$^1$}
\author{Kwan-Woo Lee$^{1,2}$}
\email{mckwan@korea.ac.kr}
\affiliation{
 $^1$Department of Applied Physics, Graduate School, Korea University, Sejong 30019, Korea\\
 $^2$Division of Display and Semiconductor Physics, Korea University, Sejong 30019, Korea}
\author{Warren E. Pickett}
\email{wepickett@ucdavis.edu}
\affiliation{Department of Physics, University of California, Davis, California 95616, USA} 
\date{\today}
\begin{abstract}
Atomic $4f$ states  have been found to be
essential players in the physical behavior of lanthanide
compounds, at the Fermi level $E_F$ as in the proposed topological 
Kondo insulator SmB$_6$,
or further away as in the magnetic superconductor system 
${\cal R}$Ni$_2$B$_2$C 
(${\cal R}$=rare earth ion)
and in Y$_{1-x}$Pr$_x$Ba$_2$Cu$_3$O$_7$, where the $4f$ shell of Pr has a 
devastating effect on superconductivity. 
In hole-doped ${\cal R}$NiO$_2$, the ${\cal R}$=Nd member is found to be
superconducting while ${\cal R}$=La is not, in spite of the calculated electronic
structures being nearly identical.
We report first principles results that indicate that the Nd $4f$ moment 
affects states at $E_F$ in infinite-layer NdNiO$_2$, 
an effect that will not occur for LaNiO$_2$.
Treating 20\% hole-doping in the virtual crystal
approach indicates that 0.15 holes empty the $\Gamma$-centered Nd-derived
electron pocket while leaving the other electron pocket unchanged; hence Ni
only absorbs 0.05 holes; the La counterpart would behave similarly.
However, coupling of $4f$ states to the electron pockets at $E_F$ 
arises through the Nd 
intra-atomic $4f-5d$ exchange coupling $K\approx 0.5$ eV
and is ferromagnetic (FM), {\it i.e.} anti-Kondo, in sign.
This interaction causes spin-disorder broadening
of the electron pockets and should be included in
models of the normal and superconducting states of \Sno. 
The Ni moments differ by 0.2$\mu_B$ for FM and antiferromagnetic
alignment (the latter are larger), reflecting some itineracy and
indicating that Heisenberg coupling of the moments may not
provide a quantitative modeling of Ni-Ni exchange coupling.
\end{abstract}
\maketitle

\section{Background and Major Issues}
The discovery of superconductivity up to T$_c$=15 K in Sr-doped
\nno~(\Sno)\cite{li2019} has re-invigorated the three decade long issue of whether 
there may be nickelates that can host cuprate-type superconductivity. 
\lno, which is isostructural with ``infinite layer'' \cco\cite{siegrist1988} 
that superconducts up to 110 K when doped,\cite{azuma1992} has been 
one of the prime candidates, but what doping is possible has never 
resulted in superconductivity.  \nno~is also isovalent with \cco, 
providing a nominal $d^9$ configuration on the open shell transition metal ion. 

Early comparison of the electronic structures of \cco~ and \nno~ identified
similarities but substantial differences,\cite{anisimov1999,lee2004}
with questions about whether concepts such as strong superexchange coupling 
of the transition metal moments, or of Zhang-Rice singlets upon 
doping,\cite{li2019}
are relevant to this nickelate.
A number of groups recently have revisited this question, with the objective 
of few-band model-building to quantify differences related to 
\scy.\cite{botana2019,k.kuroki,sawatzky,thomale,nomura2019,gao2019,m.j.han,g.zhang,
h.zhang,vishwanath,jiang2019,hu2019,werner2019,liu2019}
The popular approach has been to obtain the non-magnetic 
local density approximation 
(LDA) band structure of \nno~with the $4f^3$
electrons assigned to a non-magnetic core, then approximate the 
low energy bands with a minimal local orbital model, add a repulsive 
local interaction, and evaluate the consequences, {\it viz.} 
the pairing susceptibility.
Close attention is given to 
the geometry of the Fermi surface (FS), which strongly impacts the pairing 
susceptibility. 

Detailed analyses of the LDA electronic structure have 
been provided by Botana and Norman\cite{botana2019} for \lno~and 
by Nomura {\it et al.}\cite{nomura2019} for \nno, 
in which three non-magnetic $4f$ electrons are included in the Nd core. 
In addition to 
the roughly half-filled Ni $3d_{x^2-y^2}$ band, several groups find and 
quantify Nd $5d$ derived bands obtained earlier.\cite{anisimov1999,lee2004} 
A Nd $5d_{z^2}$ band 
drops below the Fermi level $E_F$ at $\Gamma$, and in addition a band 
usually identified as Nd 
$5d_{xy}$-derived drops below $E_F$ at the zone corner $A$=($\pi,\pi,\pi$) point 
[in units of ($\frac{1}{a},\frac{1}{a},\frac{1}{c}$)]. 
This undoped \nno~FS differs in
significant ways from that of nonmagnetic \cco~but is similar to \lno. 
X-ray absorption spectra of the O K and Ni L$_3$ edges\cite{hepting} 
do not reveal significant differences between \lno~and \nno.

There are other fundamental questions to be considered. Experimentally, 
undoped \cco~orders antiferromagnetically with a high Neel 
temperature of 442 K,\cite{mikhalev2004} while undoped \nno~and \lno~do 
not order.\cite{li2019} The underlying differences with cuprates
are several. The mean Ni $3d$
level $\epsilon_d$ is separated from the O $2p$ level $\epsilon_p$
by 4 eV in the nickelates versus only 2 eV in the cuprate.\cite{botana2019} 
Ni therefore has less 
hybridization with oxygen $2p$ orbitals than does Cu, thus should have a 
stronger tendency toward local moment magnetism than does Cu, and
indeed it does so
in LDA calculations. The superexchange coupling $J$ 
is much smaller for the nickelate,\cite{botana2019}
thus one expects a lower but nonzero Neel temperature. Finally, \lno~and \nno~are 
reported as conducting (albeit poorly), 
whereas \cco~is insulating. Lack of Ni ordering is 
a central question to address; however, both in calculations and experimentally
the nickelates are conducting. We note that magnetic order disappears in cuprates
when they are doped to become conducting.

The model treatments mentioned above would suggest that doped \lno~should be
superconducting as is doped \nno.
Since this is not the case,
we pursue the viewpoint that a more fundamental question is what features
may account for this difference between two highly similar nickelates.
La and Nd are very similar chemically, 
with tri-positive ionic radii only slightly larger for La than for Nd.
Although small size differences in lattice constants or strains can
influence states in correlated insulators, itineracy and screening
by carriers strongly reduce such effects. 

An obvious difference between these nickelates 
is that La$^{3+}$ is closed shell and nonmagnetic, whereas 
Nd$^{3+}$, with Hund's rule ground state $S=\frac{3}{2}, L=6, J=\frac{9}{2}$, 
has a (free ion) Curie-Weiss moment of nearly 4 $\mu_B$. Superconductivity
appears in \Sno~in the midst of these large disordered local moments, which should compete
with superconducting phases, directly or by introducing disorder into
the electronic structure. In several YBCO-type cuprates, however, 
replacing a nonmagnetic rare earth ion (La or Y) by a magnetic lanthanide
from Dy to Tb 
does not affect the superconductivity, because the small $4f$ orbital
is not involved in the electronic structure. 
Surprisingly, PrBa$_2$Cu$_3$O$_7$ was found to be
non-superconducting,\cite{radousky} due to antibonding coupling of the 
Pr $4f_{z(x^2-y^2)}$ orbital to the O $2p$ orbitals.\cite{ail.iim,wep.iim} Notably, the
local environment of the $4f$ ion is the same as in \nno.

Another related system is the rare earth nickel borocarbides 
${\cal R}$Ni$_2$B$_2$C where ${\cal R}$ can be one of many of the rare earth
atoms. This system displays strong interplay and competition between
the ${\cal R}$ $4f$ magnetic moment and the conduction states with
heavy Ni character, with the ${\cal R}$ $5d$ playing an important
role in coupling the local moments to the itinerant 
states.\cite{Gupta2006,wepdjs} Elemental europium at high pressure
but retaining a $f^7$ local moment, switches from metallic and 
magnetically ordered to superconducting and magnetically disordered
around 80 GPa,\cite{schilling,pi2019}
providing its own questions about the effect of $4f$ moments on 
pairing, a coupling that must proceed through the Eu $5d$ states.

The question of possible impact of the $4f$ shell has not 
been addressed in \nno. While direct $4f$-nickelate hybridization may be small,
there is an intra-atomic exchange coupling 
between the $4f$ spin and the $5d$ states around $E_F$ that, as mentioned, 
does not occur in \lno.  Since the $5d$ bands 
cross $E_F$, carriers will be affected. Specifically, with 
disordered Nd moments, the $5d$ on-site energies will be spin-split, with 
a projection along any given direction that is distributed randomly. A 
carrier in a $5d$ band will be subjected to a random potential, resulting 
in band broadening and magnetic scattering that is absent in \lno. 

The presentation is organized as follows.
The computational methods and material configurations are described in
Sec. II. In Sec. III results are presented for four assumed types of
magnetic order in \nno. The electronic bands and Fermi surfaces for
non-magnetic Ni are discussed in Sec. IV. A few results relating to the
occupied Nd $4f$ orbitals are presented in Sec. V, followed by a discussion
of hole doping in Sec. VI. Some observations about the effects of
magnetic order and relation to other $4f$-containing superconductors
are made in Sec. VII, which is followed by a brief summary in Sec. VIII.

\section{Computational Methods}
We have studied the electronic and magnetic properties of \nno~ using the
precise, all-electron, full potential linearized augmented planewave method 
as implemented in {\sc wien2k}\cite{wien2k}. The lattice constants 
$a=3.92$ \AA, $c=3.37$ \AA
~observed for superconducting \Sno\cite{li2019}
have been used, all atoms lie at $P4/mmm$ symmetry determined positions. 
Strong intra-atomic interactions on Nd, and
usually on Ni in nickelates, are modeled with the 
DFT+U method\cite{ylvisaker2010} (density functional
theory plus Hubbard $U$), which is essential to preserve the $4f$ local
moment and place the corresponding bands away from $E_F$.  
The fully anisotropic, rotationally invariant DFT+U correlation correction
functional in the fully localized limit form\cite{shick1999} implemented in {\sc wien2k}, 
was used.
Results are presented for 
  $U_f^{Nd}$=8.0 eV, $J_f^{Nd}$=1.0 eV, $U_d^{Ni}$=5.0 eV, and $J_d^{Ni}$=0.7 eV
 (and occasionally for $U_d^{Ni}$=0).
Other input parameters and a survey of results beyond those presented here
are provided in the Supplemental Material (SM).\cite{SM} 
Section VI was carried out by reducing the nuclear charge on Nd by 0.20,
and removing the same amount of electrons.

The strong local $4f$ moment requires magnetically ordered calculations.
We find that Ni prefers a moment of the order of 0.5 $\mu_B$ without any
$U^{Ni}$, which increases to $\sim$1 $\mu_B$ for the values 
we have used.  The Nd moment always
assumes the spin value of 3 $\mu_B$ characteristic of an $f^3$ Hund's
rule ion, plus minor exchange polarization of the $5d$ orbitals. 
We have studied four magnetic configurations:\\
AFM0: antiferromagnetic (AFM) Nd layers with non-magnetic Ni layers;\\
AFM1: both Ni and Nd layers have antialigned moments; \\
AFM2: ferromagnetic (FM) Nd layers and AFM Ni layers; and\\
FM: all Nd and Ni moments aligned.\\
Note that it is the AFM1 and AFM2 cases in which Ni is antiferromagnetically
ordered.
Note also that AFM0 is done with $U_d^{Ni}$=0, so its energy
cannot be compared with the other cases.

The SM provides additional information about the electronic structures.
An insulating band structure is never obtained, which is a great difference
compared with cuprates. It was shown earlier that increasing $U$ on 
Ni in NdNiO$_2$ never
leads to a Mott insulator, instead a $3d_{z^2}$ orbital becomes unoccupied and
forms a peculiar intra-atomic singlet with the $d_{x^2-y^2}$ 
orbital.\cite{lee2004}

\begin{figure}[tbp]
\centerline{\includegraphics[width=0.55\columnwidth]{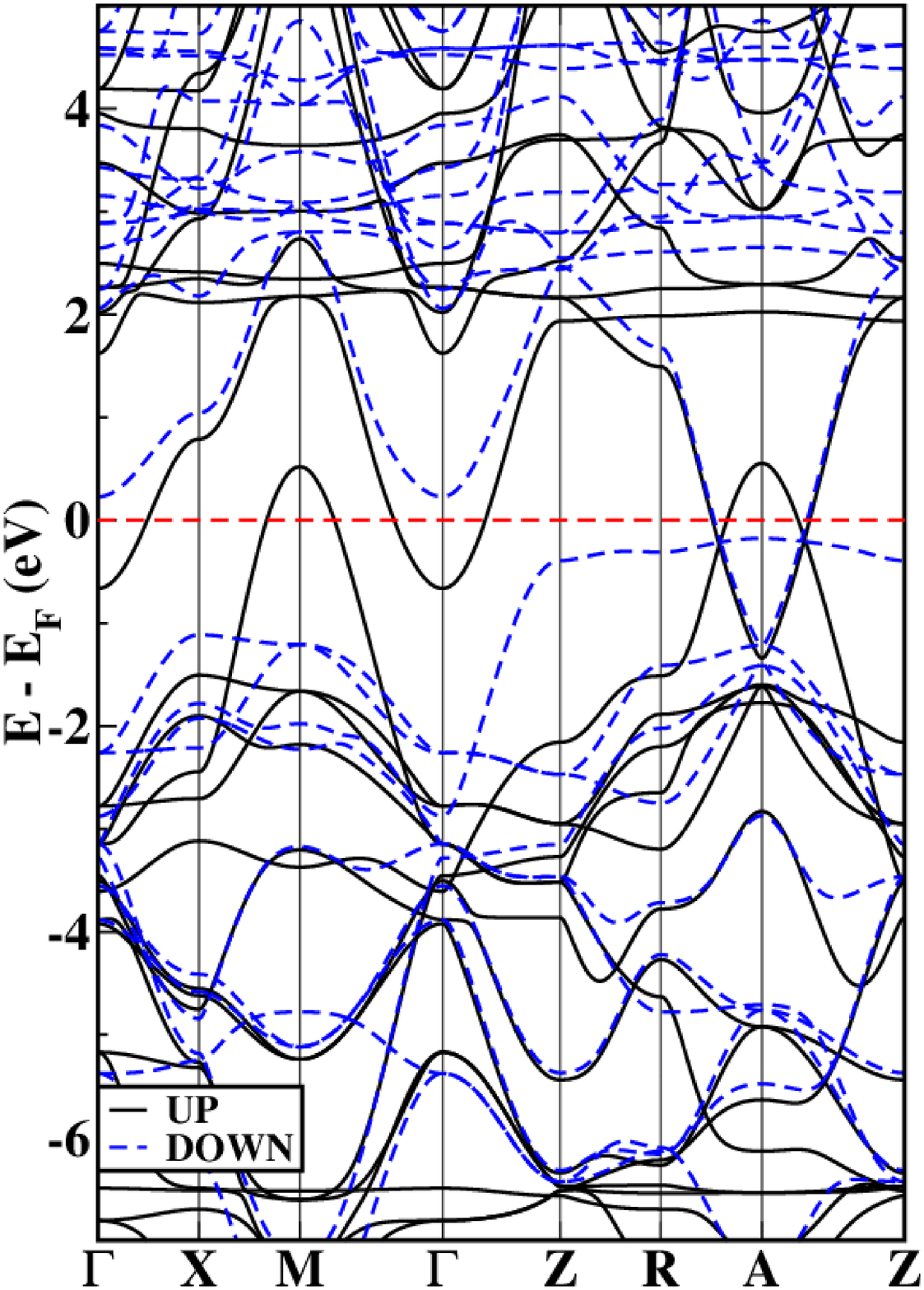}}
\vskip 2mm
\centerline{\includegraphics[width=0.7\columnwidth]{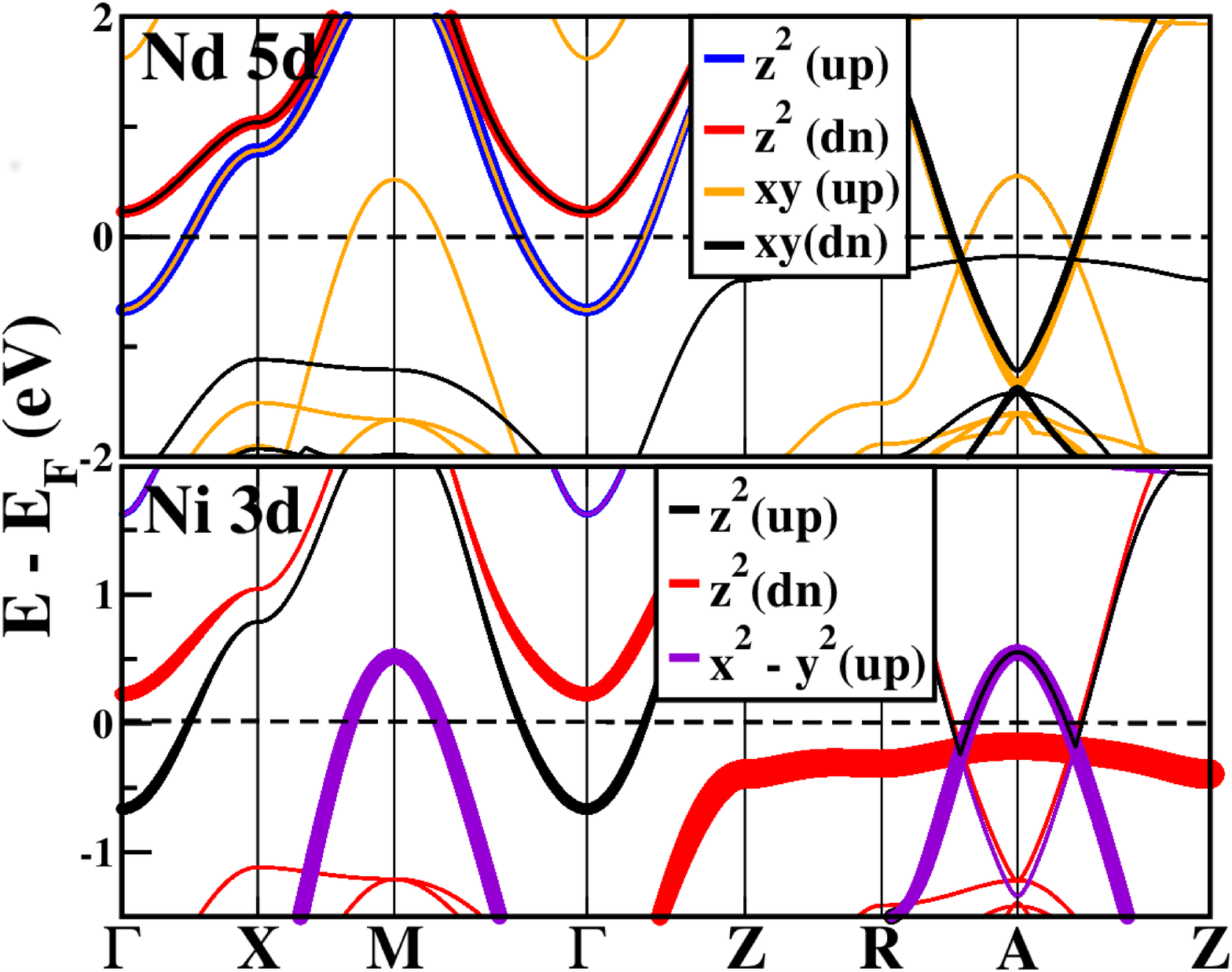}}
\caption{Top: GGA+U FM bands of \nno: solid lines denote majority,
dashed lines indicate minority, and $E_F$ is the zero of energy.
Bottom: Color fatband plot of FM \nno~ near $E_F$,
with the legend describing the color choice. Note the
Nd-derived electron pockets at $\Gamma$ and $A$, and the
Ni majority $d_{x^2-y^2}$ hole pockets at $M$ and $A$.
}
\label{fmband}
\end{figure}

\section{Magnetic order}
Although it is expected that Ni$^{1+}$ $d^9$ ions will tend to order 
in antiferromagnetic (AFM1 or AFM2) fashion through superexchange as does
\cco~and several other 
$d^9$ cuprates, we begin with FM alignment within both Ni and Nd layers,
as it enables identification of the intra-atomic exchange splitting between the
occupied Nd $4f$ and conduction $5d_{z^2}$ orbitals. 

A magnetic Ni ion leads to a 120 meV energy gain compared to non-magnetic.
This exaggerated tendency toward magnetic ordering and the magnitude of
the magnetic moment are a known deficiency of (semi)local density
functional methods, with much of the problem attributed to the lack
of effects of spin fluctuations in the functional.

The energies can be compared 
for those cases with magnetic Ni (AFM1, AFM2, FM) for which
the same functional is used. 
The energy of AFM2 (FM Nd) is
slightly lower than that of AFM1 (AFM Nd), by 7 meV for $U^{Ni}$=0 and
by only 1 meV (at the edge of computational precision) with $U^{Ni}$=5 eV,
reflecting a slight tendency toward FM ordering of the $4f$ moments.
The energy difference between FM and AFM2 provides the difference between
AFM and FM Ni moment alignment; AFM is favored by 116 (25) meV/Ni in
GGA(+U). This energy difference contains information about Ni-Ni in-plane
exchange coupling, and would give a value of the nearest neighbor coupling
if more distant exchange couplings were negligible. 
Liu {\it et al.} have derived values of
coupling for a few neighbors, concluding that the values depend strongly
on the value of $U^{Ni}$ that is assumed.\cite{liu2019}
 
For the FM bands shown in Fig.~\ref{fmband} the Ni and Nd moments are 
aligned into an overall FM structure. The Ni $d_{x^2-y^2}$ bands that
give rise to the large FS are spin-split by 2 eV, 
reflecting the Ni spin moment of order 1$\mu_B$.
As noted earlier\cite{anisimov1999,lee2004} and recently 
by several others, Nd $5d$-derived bands dip below $E_F$ at 
the zone center $\Gamma$ and 
at the zone corner $A$, with bonding $5d_{z^2}$ and antibonding $5d_{xy}$ 
character respectively.\cite{nomura2019} There is mixing with 
Ni $d_{z^2}$ in both of these electron pockets, especially the latter. 
This mixing is due to the surprise that
the minority (but occupied) Ni $d_{z^2}$ antibonding band at $k_z=\pi$ 
(see the $Z-R-A-Z$ lines) 
lies within 0.2--0.4 eV of $E_F$.

As shown in Fig.~\ref{fmband}, the $5d_{z^2}$ band energies
relative to $E_F$ of up (down) bands are --0.7 (+0.3) eV at 
$\Gamma$, giving an
intra-atomic coupling $H^{Nd}=K{\hat e_{4f}}\cdot {\hat e_{5d}}$ 
with $K$=0.5 eV,
in terms of the orientations of the $4f$ and $5d$ spins. 
For the antibonding $5d_{xy}$ orbital the eigenvalues at $A$ lie at
--1.2 eV with a splitting of less than 0.1 eV, reflecting
the participation of Ni $3d$ orbitals and the out-of-phase
character of states at the $A$ point. 
Note that this coupling is Hund's rule alignment of
$4f$ and $5d$ states and therefore anti-Kondo coupling of the local
moment to the conduction band, as opposed
to some suggested Kondo modeling of \Sno.\cite{anisimov1999,hepting} 
Such anti-Kondo coupling
has been studied in other lanthanide compounds.\cite{kunes2004}
We return below to the $K\approx 0.5$ eV intra-atomic exchange 
coupling of Nd $5d$ states. 

\begin{figure}[tbp]
\centerline{\includegraphics[width=0.8\columnwidth]{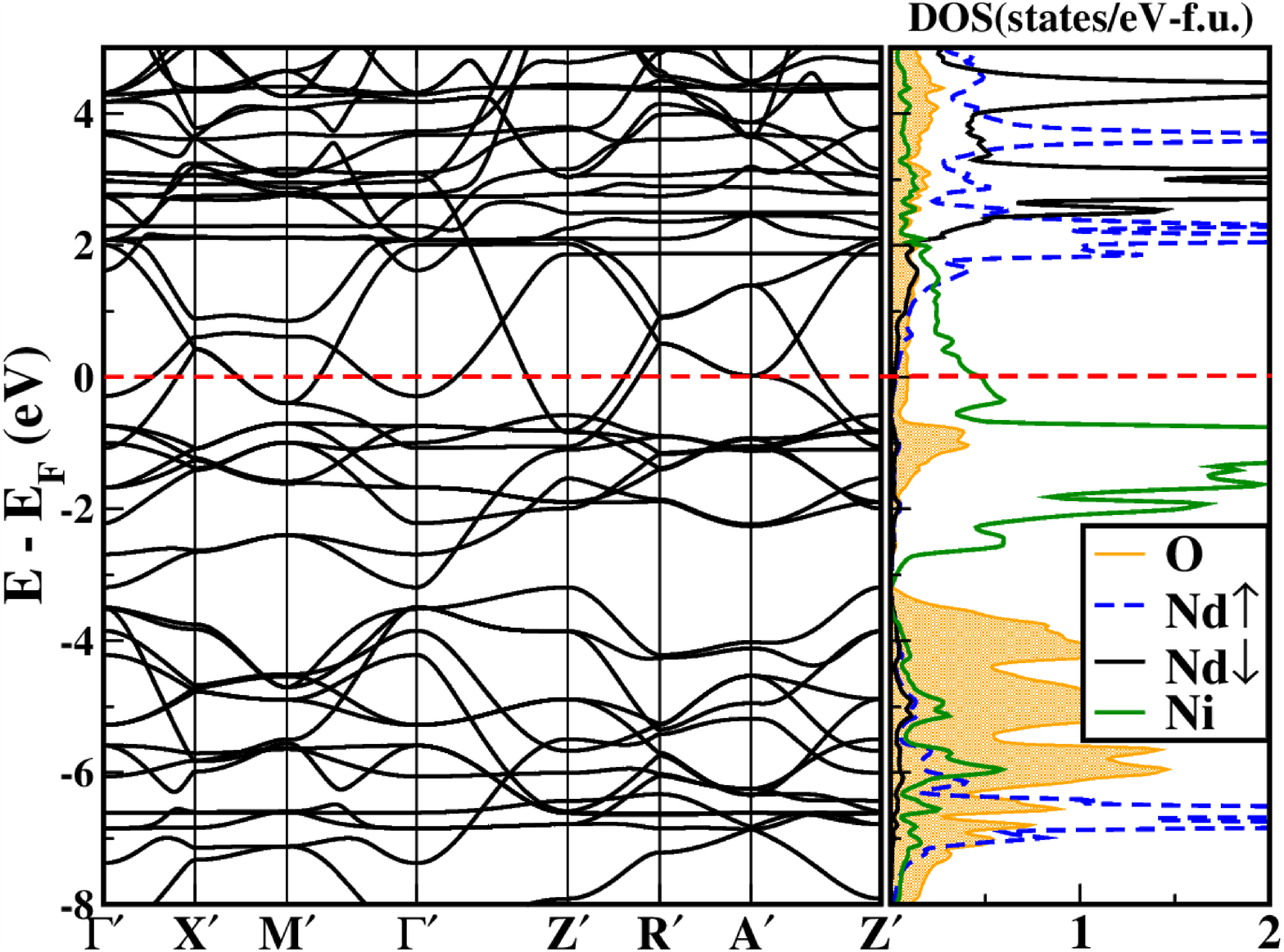}}
\vskip 5mm
\centerline{\includegraphics[width=0.8\columnwidth]{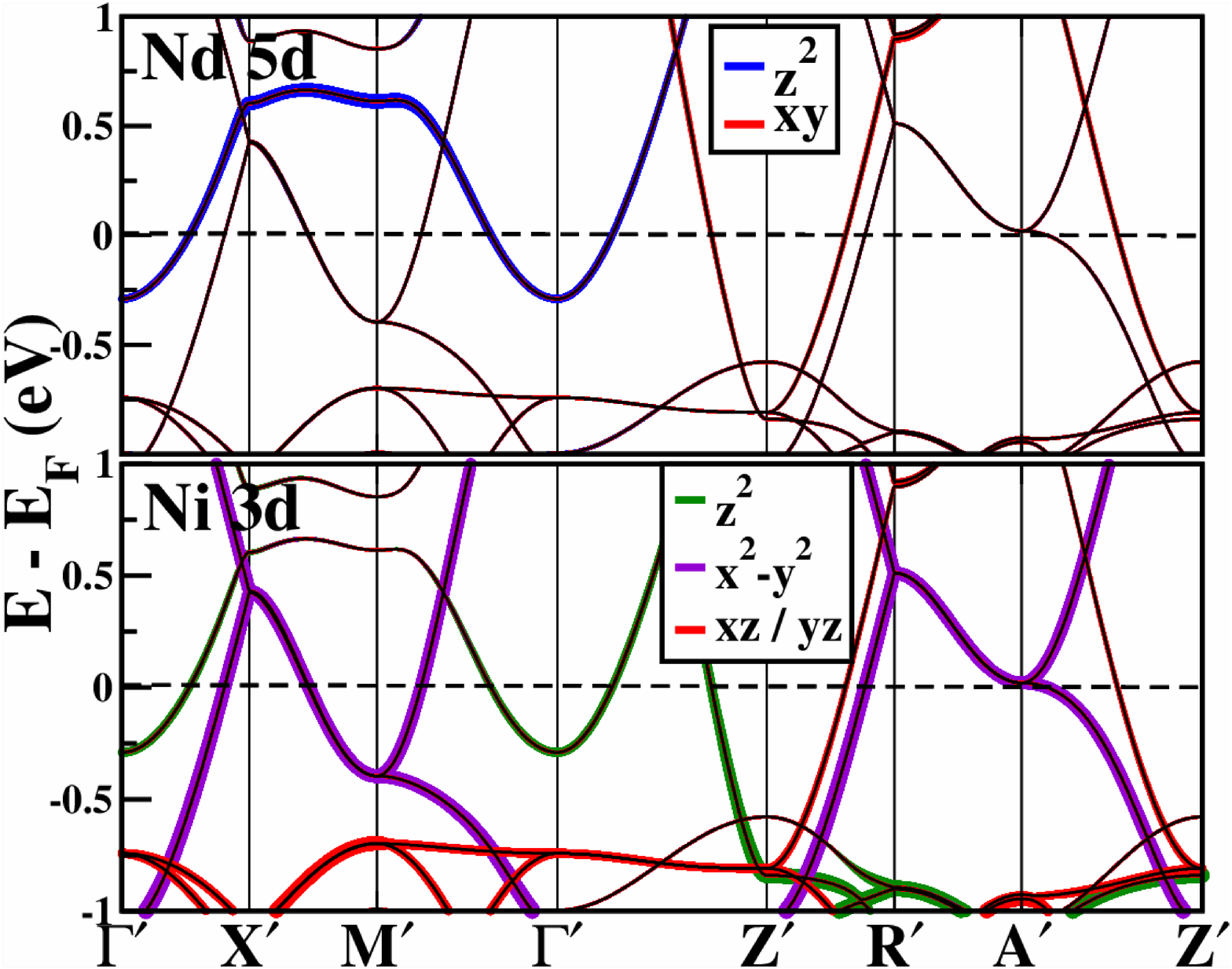}}
\caption{Top: Bands of \nno, with AFM Nd moments and non-magnetic
Ni layer, plotted along lines in the AFM zone.
Bottom: Fatbands indicating Nd $5d$ character (above) and Ni $3d$
weight (below), with orbital character as noted. In terms of the usual
NiO$_2$ sublattice, 
the $M$($A$) point is folded back to $\Gamma'$($Z'$). 
The other symmetry points are $X'=(\pi/2,\pi/2,0)$ and $M'=(\pi,0,0)$, 
while $R'$ and $A'$ lie above these points by $(0,0,\pi)$ respectively.
}
\label{gga_u}
\end{figure}

\section{Nonmagnetic N\lowercase{i} and Fermi surfaces}
Low energy models of oxides
including superconducting possibilities are based on
non-magnetic Ni, with an on-site repulsion added to provide correlated
electron behavior and a potential pairing mechanism.
We have obtained a DFT+U solution with non-magnetic Ni 
($U_d^{Ni}$=0, $U_f^{Nd}$=8 eV),
our case AFM0.  The Nd $4f$ moment that
must remain fully spin-polarized has AFM alignment, 
so the Brillouin zone is
correspondingly folded back. 
The Ni moment for AFM
alignment is 1.2$\mu_B$, compared to 1.0$\mu_B$ for FM alignment.
For $U$=0 (just GGA) the calculated moments are 0.52$\mu_B$ (AFM) and
0.35$\mu_B$ (FM); {\it i.e.} there is about 0.2$\mu_B$ difference in
the moments, and the associated Hund's energy would be an 
important contribution to the lower energies for AFM alignment.

The bands, density of states (DOS), 
and fatbands near $E_F$ for AFM0 alignment are
shown in Fig.~\ref{gga_u}, with symmetry points in that folded
tetragonal zone now
designated with primes.  Gaps at the zone boundaries indicate effects of the
Nd AFM order, {\it e.g.} a gap of 0.3 eV can be seen in the Nd $5d$ band 
around 0.8 eV at $X'$.
As for the FM order above, this splitting vanishes for the corresponding
antibonding band at $R'$. The nearby folded Ni $3d$ band displays no gap,
reflecting canceling mixing with the eight neighboring Nd sites. 

\begin{figure}[!htbp]
\centerline{\includegraphics[width=0.95\columnwidth]{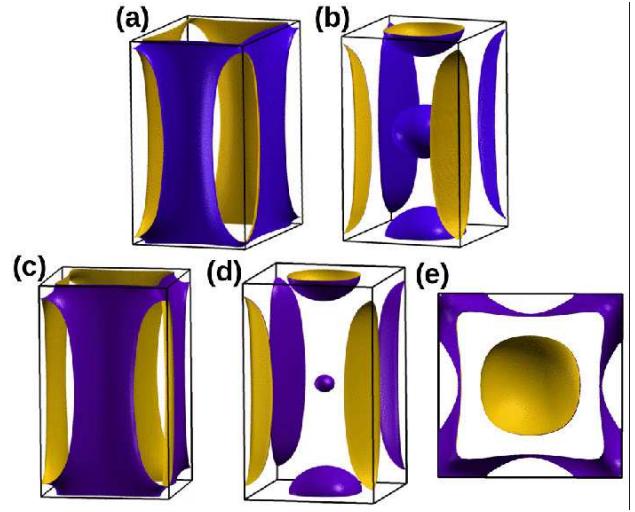}}
\caption{Fermi surfaces for case AFM0 (non-magnetic Ni),
in the full zone corresponding to AFM $4f$ alignment. Surfaces
have been separated into left and right sub-panels for clarity.
(a),(b) Undoped \nno.
The spheres are (i) the Nd $5d$ electron pocket at $\Gamma'=\Gamma$,
at the center of the figure, and (ii) the
other electron pocket at $A$ folded back to $Z'$ (see text).
(c),(d): Surfaces for VCA \Sno, $x$=0.20 hole doping. The $\Gamma$-centered
electron pocket is essentially emptied, while the $Z'$-centered pocket
is almost unchanged. (e) Top view of the \Sno~surfaces, showing
the amount of $k_z$ dispersion.
}
\label{fermisurfaces}
\end{figure}

The bands crossing $E_F$ giving the FSs are those that are fit to few band models. 
The corresponding FSs are shown in Fig.~\ref{fermisurfaces}.
The Nd $5d_{z^2}$ related electron pocket Pk1 at $\Gamma$ has attracted attention,
with the electron pocket Pk2 at $A$=($\pi,\pi,\pi$) [$Z'$ in Fig.~\ref{gga_u}]
also receiving notice. Pk2 has strong Ni $3d_{z^2}$ character as well as some
$3d_{xz,yz}$ influence, considerably stronger than the Nd $5d_{xy}$ component that
has been discussed by several 
groups.\cite{lee2004,botana2019,thomale,gao2019,nomura2019,k.kuroki} 
Due to $k_z$ dispersion, the large Ni $d_{x^2-y^2}$ FS appears in this folded
zone as a large banana centered at $M'$, and strongly fluted cylinders
also centered at $M'$ and connected by tiny necks at the corner $R'$ points.
These two surfaces are degenerate along the zone faces reflecting simple folding
back into the AFM zone, and correspond to $M$-centered hole barrels in the primitive
zone that are familiar from cuprate physics. These barrels have surprisingly
large $k_z$ dispersion, lessening the relevance of two dimensional physics.

\section{Occupied $4f$ orbitals}
We now consider briefly the Nd $4f$ contributions to the electronic
and magnetic structure. The $S=\frac{3}{2}$ spin
configuration is strongly enforced by Hund's first rule, and the energy
difference between FM and AFM order of the Nd moments is very small; $\sim 7$ 
meV/Nd ion for Ni treated in GGA, and only 1 meV/Nd ion for Ni treated in GGA+U.
The three occupied bands lie at --7 eV
with some small dispersion due to the direct overlap of $4f$
and O $2p$ orbitals, which is small but provides crystal field splitting
of the $4f$ orbitals. The occupied states that we obtained
are combinations of $m_{\ell}=
-3, -2, +1$, each with lesser admixtures of $+1, +2, -3$ states,
respectively, with total orbital moment $m_{orb}$=4.47 $\mu_B$.
When spin-orbit coupling is included, 
the occupied orbitals are those with $j=\frac{5}{2}, 
m_j=-\frac{5}{2},-\frac{3}{2},+\frac{1}{2}$. 
(See SM for further information on these occupations.)
The four unoccupied $4f$ bands are centered 3.5 eV
above $E_F$ but are spread by the anisotropy of the orbitals
and mixing with Nd $5d$ states over a range of 3 eV.

\section{Hole doping}
\nno~becomes superconducting upon hole doping by 0.2 hole/f.u.\cite{li2019} With
the Fermi level density of states N(0)=0.70/(eV cell), 
the drop in $E_F$ (in rigid band) is 0.29 eV, which
is enough to empty the Pk1 pocket at $\Gamma$ to 0.4 eV, and to reduce
the carrier density of Pk2 but leaving substantial Ni carriers. Rigid
band treatment may be deceptive, however, due to Fermi level charge on both
Ni and Nd atoms. We applied the virtual crystal approximation\cite{vca} (VCA) to obtain
a more realistic effect of doping. 

\begin{figure}[!htbp]
\centerline{\includegraphics[width=0.55\columnwidth]{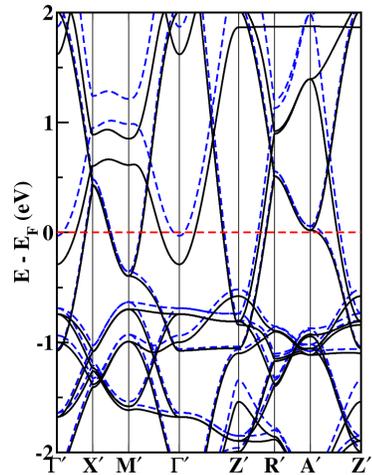}}
\caption{Band structures of \Sno~(dashed lines) compared to those of \nno~ 
 (solid lines), near $E_F$. 
Here $U^{Ni}_d=0$; Ni is non-magnetic. The 
bands are plotted in the Nd AFM zone. Note that hole doping removes electrons
  primarily from the $\Gamma$-centered Nd $5f$ electron pocket; the
$Z'$ ($A$ in primitive zone) pocket is nearly unchanged.
}
\label{virtualxtal}
\end{figure}

Fermi surfaces Pk1 and Pk2 of undoped [Figs.~\ref{fermisurfaces}(a) and \ref{fermisurfaces}b)] 
and $x$=0.20 doped \nno~ [Figs.~\ref{fermisurfaces}(c)-\ref{fermisurfaces}(e)]
reveal important non-rigid band behavior, and the corresponding bands provided in
Fig.~\ref{virtualxtal} indicate the reason. The Ni $d$ bands are shifted only
slightly ($\sim 0.1$ eV) with respect to $E_F$ by doping, as is the band giving the
$A$-centered pocket (at $Z'$ in this figure). The Nd $5d$ band at $\Gamma$ is, however, 
shifted by 0.3 eV and effectively emptied of its (undoped) 0.15 electrons.
These VCA bands are similar to the VCA bands of 
Sakakibara {\it et al.}\cite{k.kuroki} for La$_{0.8}$Ba$_{0.2}$NiO$_2$,
a difference being that their La $5d$ band
is 0.2 eV higher than our Nd $5d$ band.

The band giving rise to Pk2, often referred to as Nd $5d_{xy}$, is also
heavily Ni in character (shown above, and in the SM), hence little affected as
are the other Ni $3d$ bands. The result is that $x=0.20$ doping changes the Ni
$3d$ charge by only 0.05 holes. The disappearance of pocket Pk1 and its associated
disorder scattering may contribute to the drop in resistivity by a 
factor of 2--3 upon doping.\cite{li2019}

\section{Implication of spin disorder}
The exchange splitting of the Nd $5d$ band at $\Gamma$ (near $E_F$)
reflects the intra-atomic $4f-5d$ exchange on Nd of $K\approx 0.5$ eV. 
The $S=\frac{3}{2}$ spin
moment on Nd drives this exchange splitting, which will follow the orientation
of the $4f$ moment, which is disordered. With the very weak $4f-4f$
exchange coupling noted in Sec. III, the Nd moments will be randomly oriented and
quasi-static on an electronic time scale.

In spite of the disorder broadening
on the electron Fermi pockets and resulting strong spin scattering
driven by stable, essentially classical but disordered $4f$ 
moments, superconductivity
survives in \Sno. This situation bears some similarity to that of 
elemental Eu with its $4f^7$
moment: pressure kills magnetic order around 82 GPa but not the local moment,
and there is a first order electronic but isostructural transformation to a
superconducting phase at 1.7 K.\cite{schilling} 
The exchange coupling
from the classical local moment that should be disruptive to Cooper pairing
in a specific spin configuration (singlet  or triplet) may actually be
contributing to the pairing
in Eu.\cite{pi2019} The exchange coupling in \nno~is not as pervasive --
every site in Eu has a $4f$ moment and the partially occupied
conduction band is entirely $5d$ --
but some of the physics may be related.

A related, and commonly modeled, spin-scattering  process will occur on 
the (locally) magnetic Ni ion (as treated for the Cu ion in cuprates), 
an effect that can be modeled by any of a variety of spin 
fluctuation formalisms applicable to transition metal oxides.  
The resulting strong scattering in the electron pockets is 
a potential contributor to the resistivity of ``conducting" \nno, 
where the resistivity $\rho \sim  1 m\Omega$cm varies 
little between 300 K and the 
lowest temperature measured. The inferred mean free path in
a Fermi liquid inspired 
Bloch-Boltzmann treatment would be
atomic size, meaning incoherent transport and a washed out FS.
A dynamical mean field approximation treatment 
of Ni $3d$ fluctuations\cite{m.j.han}
in LaNiO$_2$ found little shifting and broadening of bands around $E_F$,
and no mechanism for the high resistivity.
The large residual resistivity may be the result of the topotactic
synthesis technique, with indications to be learned between the
similarities and the distinctions between the doped thin
films and doped bulk materials,\cite{Li-bulk} which also have
similarly high residual resistivity.

\section{Summary}
In this work we have focused on a primary distinction between
\nno~ and \lno. When the $4f$ moment on Nd is relegated to the
core in nonmagnetic fashion, the electronic structures are 
nearly identical.
Yet the Nd compound becomes superconducting while the sister La
compound remains non-superconducting,
when 20\% hole-doped. We find that the strong Nd
$4f$ moment is intra-atomic exchange-coupled to the Nd $5d$ orbitals
by a coupling of roughly $K$=0.5 eV, giving anti-Kondo coupling between the 
Nd local moments and the lower conduction band. 
The disordered $4f$ moments will give rise to
a broadening of $5d$ bands of the order of $K$, affecting transport
and perhaps becoming implicated in pairing, either as a participant
or a disruptive agent.

The calculated Ni moments are larger by $0.2\mu_B$
for antiferromagnetic alignment than for ferromagnetic alignment.
The associated Hund's rule energy provides a tendency toward AFM
alignment that lies beyond the usual Heisenberg exchange coupling picture.
An important feature of this system is that the $\Gamma$-centered
Nd $5d$ Fermi surface pocket accepts most of the doped holes,
leaving Ni $3d$ charge changed only by 0.05 holes for 20\% doping.
The Pk2 spherical Fermi surface pocket that remains intact after doping,
and rather strong $k_z$ fluting of the Ni Fermi surfaces, suggests
the importance of $k_z$ dispersion (three dimensionality). 
Our results indicate that several aspects of the Nd ion arising from
the open $4f$ shell require attention for understanding both
normal and superconducting state properties.  

\vskip 4mm
{\bf Note Added}: 
A useful reference has just appeared on the APS Condensed Matter Physics Journal Club postings. \cite{journalclub}.

\vskip 8mm

\section{Acknowledgments}
We acknowledge informative discussions with A. S. Botana, J. Sun, M. J. Han,
E.-G. Moon, D. Lee, and J. Son,
and communication with S. Raghu.
M.Y.C. and K.W.L were supported by National Research Foundation of Korea
Grants No. NRF-2019R1A2C1009588. 
W.E.P. was supported by NSF Grant No. DMR 1607139.

\newpage
\section{Supplemental Material}
{\bf Calculation Methods}\\
Our calculations with {\sc wien2k}  were based on the generalized gradient approximation (GGA)\cite{gga}.
The effects of correlation were treated with the GGA+U approach in the fully localized limit.
In the GGA+U calculations presented here, 
we focused on the results of the Hubbard $U=8 (5)$ eV  and Hund's coupling $J=1 (0.7)$ eV 
for the Nd $4f$ (Ni $3d$), except where noted.
Our conclusions weren't affected by varying the value of $U$ by 1--2 eV.

In {\sc wien2k}, the basis size was determined by $R_{mt}K_{max}=8$; checking $R_{mt}K_{max}$=9
gave no noticeable difference.
The APW radii were chosen as Nd 2.50, Ni 1.98, and O 1.70, in atomic units.
Convergence was checked with dense $k-$meshes up to $29\times 29\times 34$ to treat the 
highly localized $4f$ orbitals carefully.\\

{\bf Fermi surface of Nd$_{0.8}$Sr$_{0.2}$NiO$_2$ in rigid band approximation}\\

\begin{figure}[htbp!]
{\resizebox{7cm}{4cm}{\includegraphics{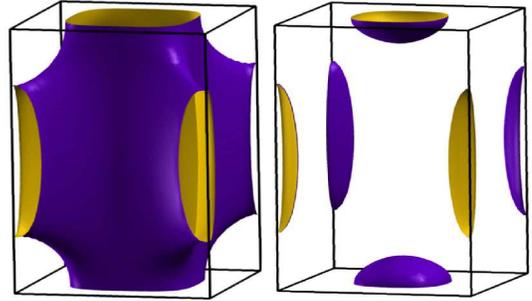}}}
\caption{Fermi surfaces of $x$=0.20 Sr hole doped \nno, in the rigid band
approximation, for case AFM0 (non-magnetic Ni) as defined in the main text.
The region shown is the folded zone corresponding to AFM $4f$ alignment. Surfaces
have been separated into left and right sub-panels for clarity. These
surfaces can be compared with the bottom panel of Fig. 3 in the main text, which provides the
surfaces in the virtual crystal approximation.
The $\Gamma$-centered
electron pocket has been emptied, while the $k_z$ dispersion is
increased substantially. (The undoped case is also given in the top panel of Fig. 3 in the main text.)
}
\label{FS}
\end{figure}


{\bf GGA+U electronic structure of ferromagnetic state}\\
We provide here and below fatband plots indicating band character for all
Ni $3d$ and Nd $5d$ orbitals, for both FM and AFM1 magnetic
alignments (fully aligned moments; both Ni and Nd moments antialigned). 
The differences, also compared to the information
in the main text for non-magnetic Ni, indicate how the various
characters are shifted by relative magnetic alignment. Although ordering does not
occur down to the lowest temperature measured, the differences give an indication
of the effects of spin fluctuations. Some of the
points to notice are included in the captions.


\begin{figure*}[htbp]
\vskip 8mm
{\resizebox{13cm}{8cm}{\includegraphics{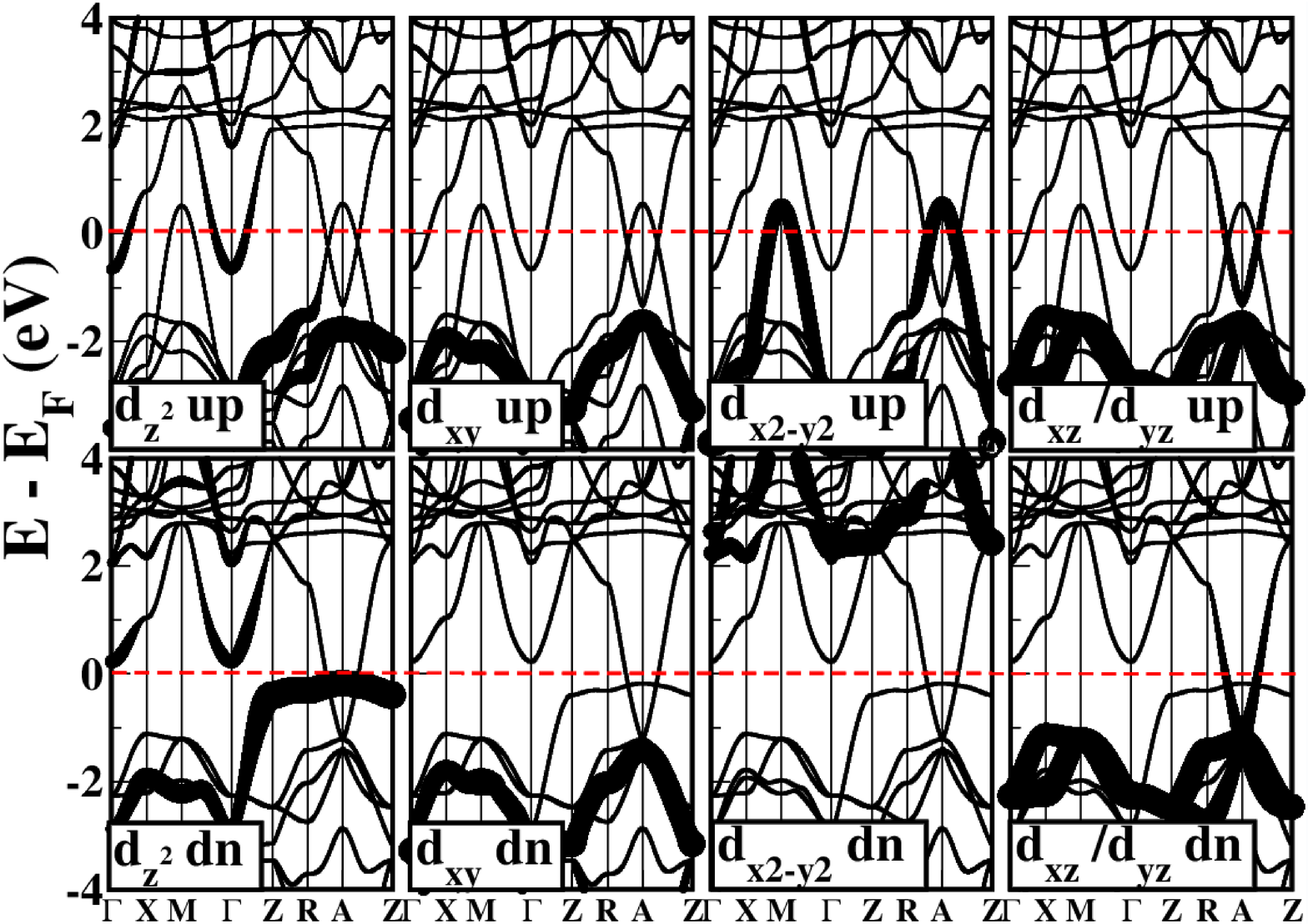}}}
\vskip 6mm
{\resizebox{13cm}{8cm}{\includegraphics{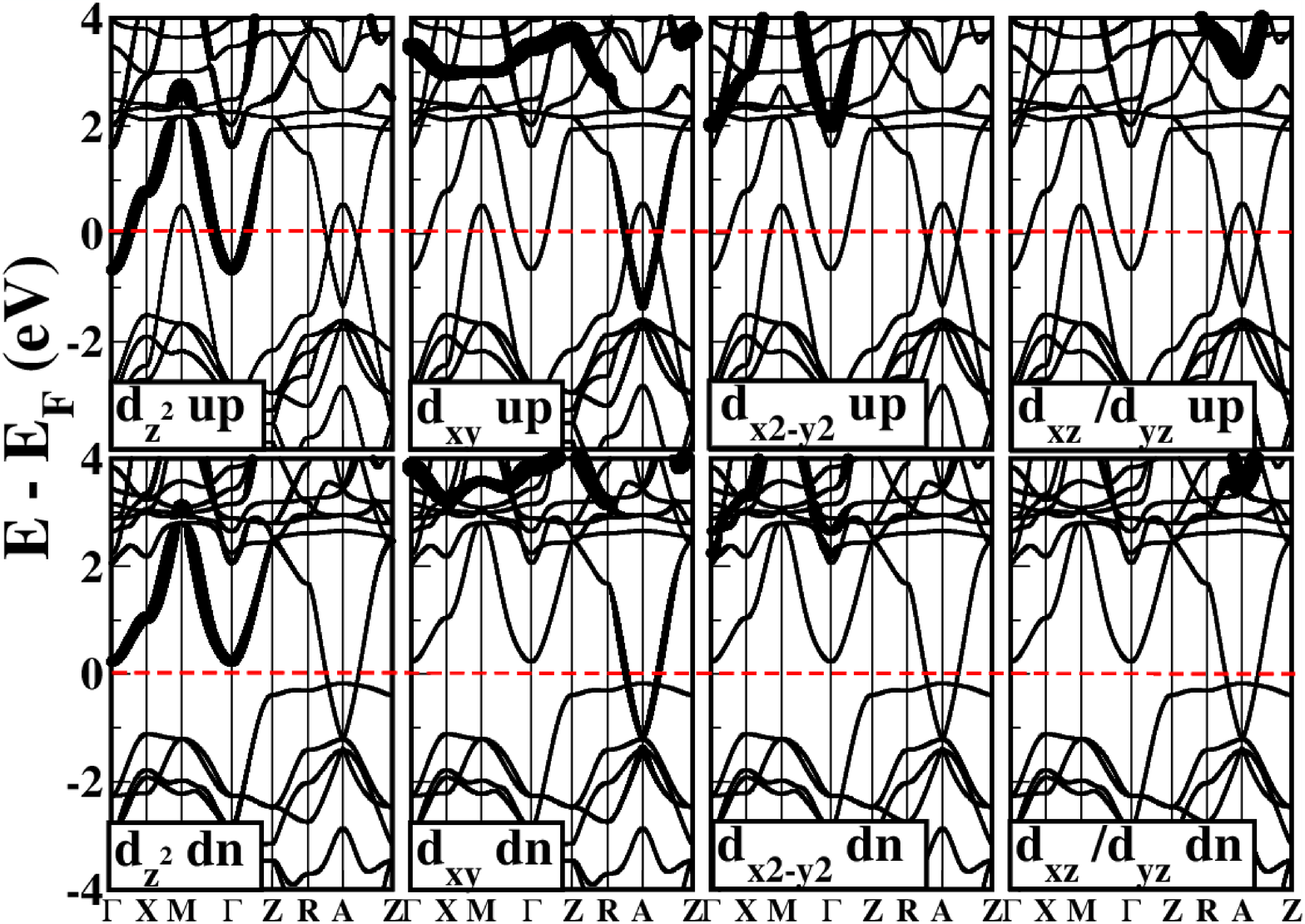}}}
\caption{Top: Fatband plots for all Ni $3d$ orbitals for FM alignment, with
the GGA+U functional (see main text and section above for 
methods and values of interactions). 
Majority (up) bands are given in the upper panels, 
minority (down) in the lower panels.
Note that there is some Ni $d_{z^2}$ character for both spin 
directions in the $\Gamma$ centered electron pocket. There is 
very strong Ni $d_{z^2}$ character almost at E$_F$ in the
minority bands along the zone edge $Z-R-A-Z$.\\
Bottom: Fatband plots for all Nd $5d$ orbitals
corresponding to those of Ni $3d$, above. 
Majority (up) bands are in the upper panels,
minority (down) bands are in the lower panels. Strong Nd $5d$
character ($d_{z^2}$) is confined to the electron pockets
Pk1 at $\Gamma$.
There is lesser Nd contribution ($d_{xy}$) to the electron pocket
Pk2 at $A$.  
}
\label{fmfat_3d}
\end{figure*}

\begin{figure*}[htbp]
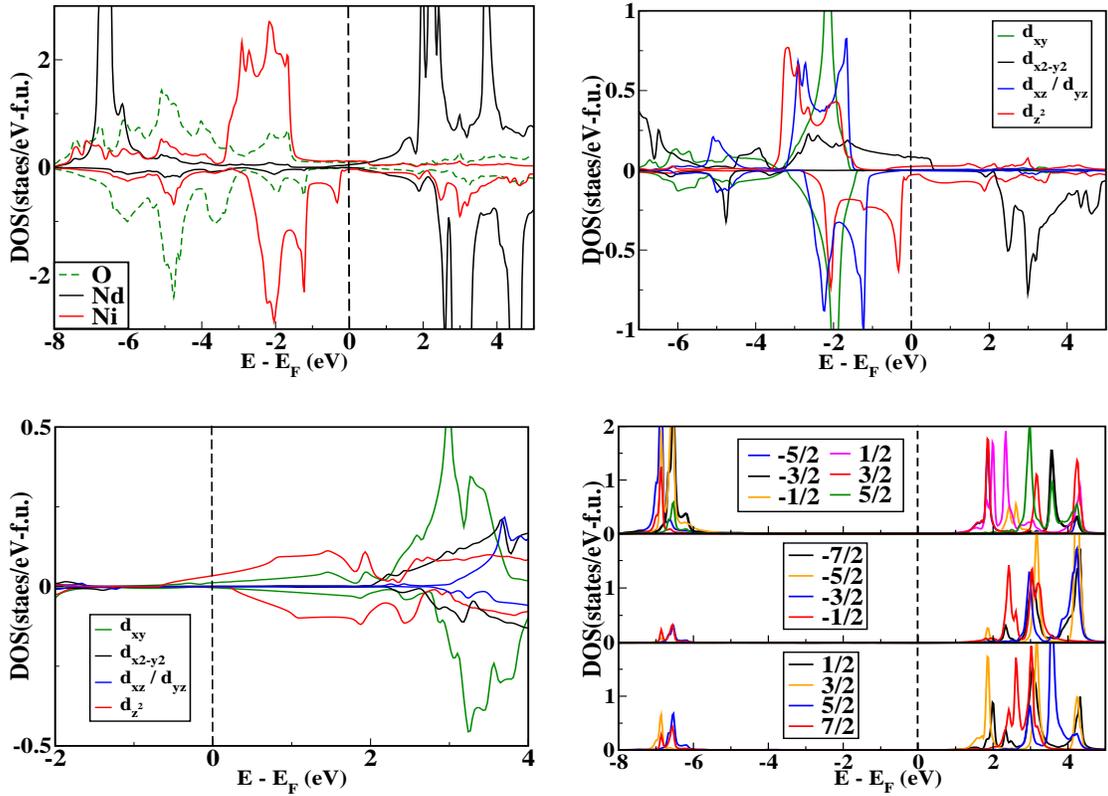

{\resizebox{7cm}{5cm}{\includegraphics{FigS3a.eps}}}
\hskip 5mm
{\resizebox{7cm}{5cm}{\includegraphics{FigS3b.eps}}}
\vskip 5mm
{\resizebox{7cm}{5cm}{\includegraphics{FigS3c.eps}}}
\hskip 5mm
{\resizebox{7cm}{5cm}{\includegraphics{FigS3d.eps}}}
\caption{FM projected densities of states (PDOSs) in GGA+U.
Top left: atom PDOS. Top right: Ni $3d$ orbital PDOS. Bottom left: Nd $5d$ 
orbital PDOS.
 Bottom right: FM orbital-resolved DOSs of Nd $4f$ orbitals in the $|j,m_j>$ basis, 
using GGA+U+SOC.
The occupied orbitals are mostly $|5/2,-5/2>$, $|5/2,-3/2>$, and $|5/2,-1/2>$ characters.
Minor mixing with $|5/2,5/2>$, $|5/2,3/2>$ orbitals lead to an orbital moment of --2.6 $\mu_B$.
}
\label{fm_pdos}
\end{figure*}

The  spin-resolved fatband plots of separate Ni $3d$ and Nd $5d$ orbitals
in the range of --4 eV to 4 eV for FM alignment of all moments
are displayed in Fig. \ref{fmfat_3d}.
Some aspects to note follow.
The electron pocket band at the $\Gamma$ point (fully spin polarized for this
FM alignment) has strong Nd $5d_{z^2}$ character as noted by others,
but it is mixed with significant Ni $3d_{z^2}$ character near the $\Gamma$ point.
For the electron pocket at $A=(\pi,\pi,\pi)$ there is admixture of
Ni $3d_{xz}/3d_{yz}$ character with the Nd $5d_{xy}$ orbitals.
The hole bands at the $M$ and $A$ points have the primary
$d_{x^2-y^2}$ character as occurs in cuprates.

Figure~\ref{fm_pdos} displays the GGA+U orbital- and atom-projected
densities of states (PDOSs) for FM alignment.
 The Nd $4f$-orbital resolved DOSs obtained from GGA+U+SOC are given
in the bottom right panel of Fig. \ref{fm_pdos}.
The main effect of SOC is to produce, via introduced anticrossings, 
individual flat bands giving rise to separated DOS peaks for each
orbital $|j.m_j>$.\\



{\bf GGA+U electronic structures of antiferromagnetic states}\\

To consider antiferromagnetic (AFM) states, a $\sqrt{2}\times\sqrt{2}$ 
supercell is necessary.
As mentioned in the main text,
we considered three AFM states and a fully aligned configuration:\\
* AFM0: nonmagnetic Ni and AFM ordered Nd  \\
* AFM1: both AFM ordered Ni and Nd ions \\  
* AFM2: AFM ordered Ni and FM ordered Nd\\
*  FM:  both sublattices completely aligned.\\
Here we provide additional information of AFM1 and AFM2, both of which
have AFM Ni layers.
For the AFM0 state, see the main text.

In GGA+U, similar to the FM cases, the magnitude of the local spin moments are  
3 $\mu_B$ for Nd and around 1.2 $\mu_B$ for Ni.
In AFM2, the FM ordered Nd moment induces a very small difference
between up and down Ni moments of 0.03 $\mu_B$,
indicating the weakness but non-vanishing of the magnetic 
Nd-Ni coupling as noted by others.

\begin{figure*}[htbp]
{\resizebox{8cm}{5cm}{\includegraphics{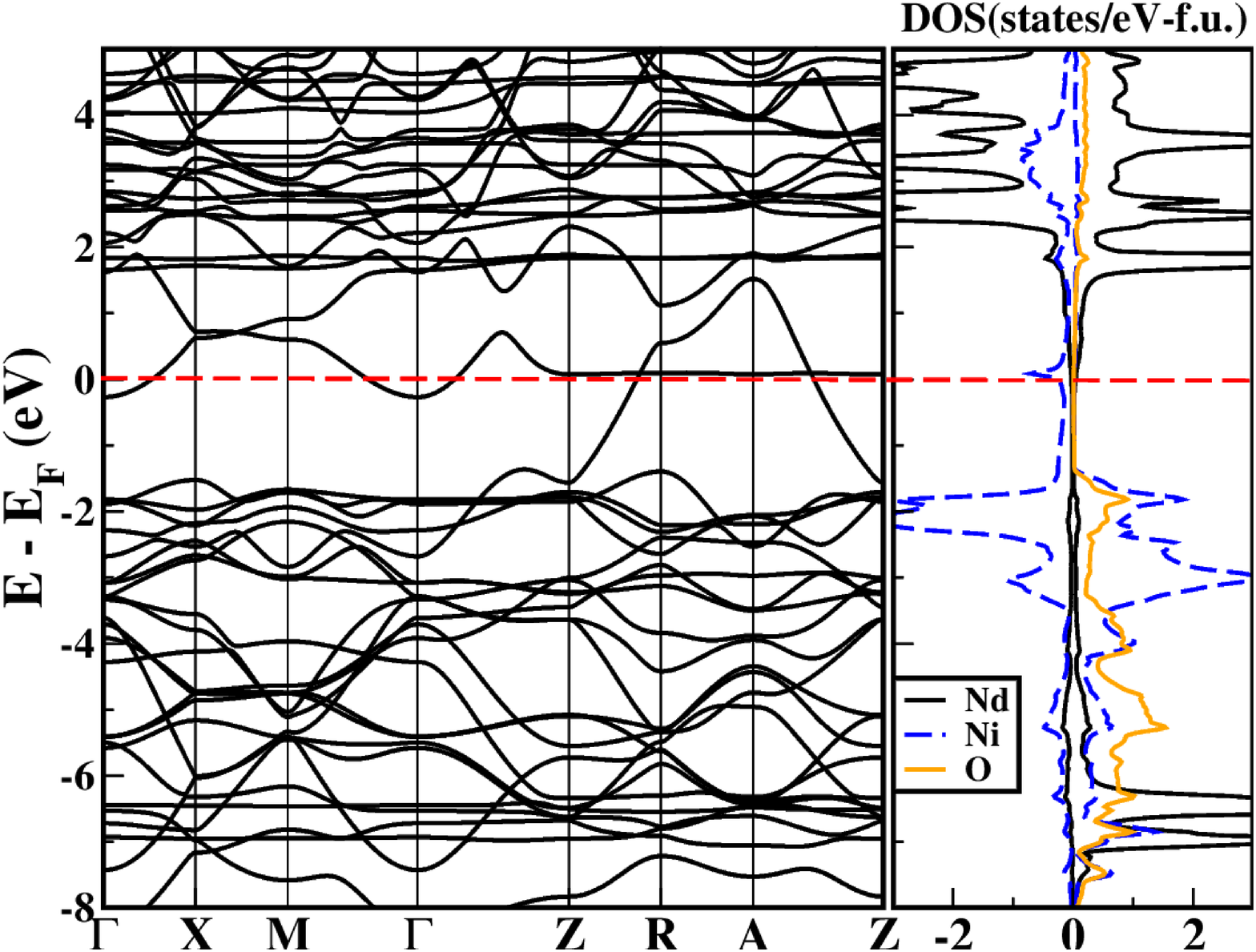}}}
{\resizebox{8cm}{5cm}{\includegraphics{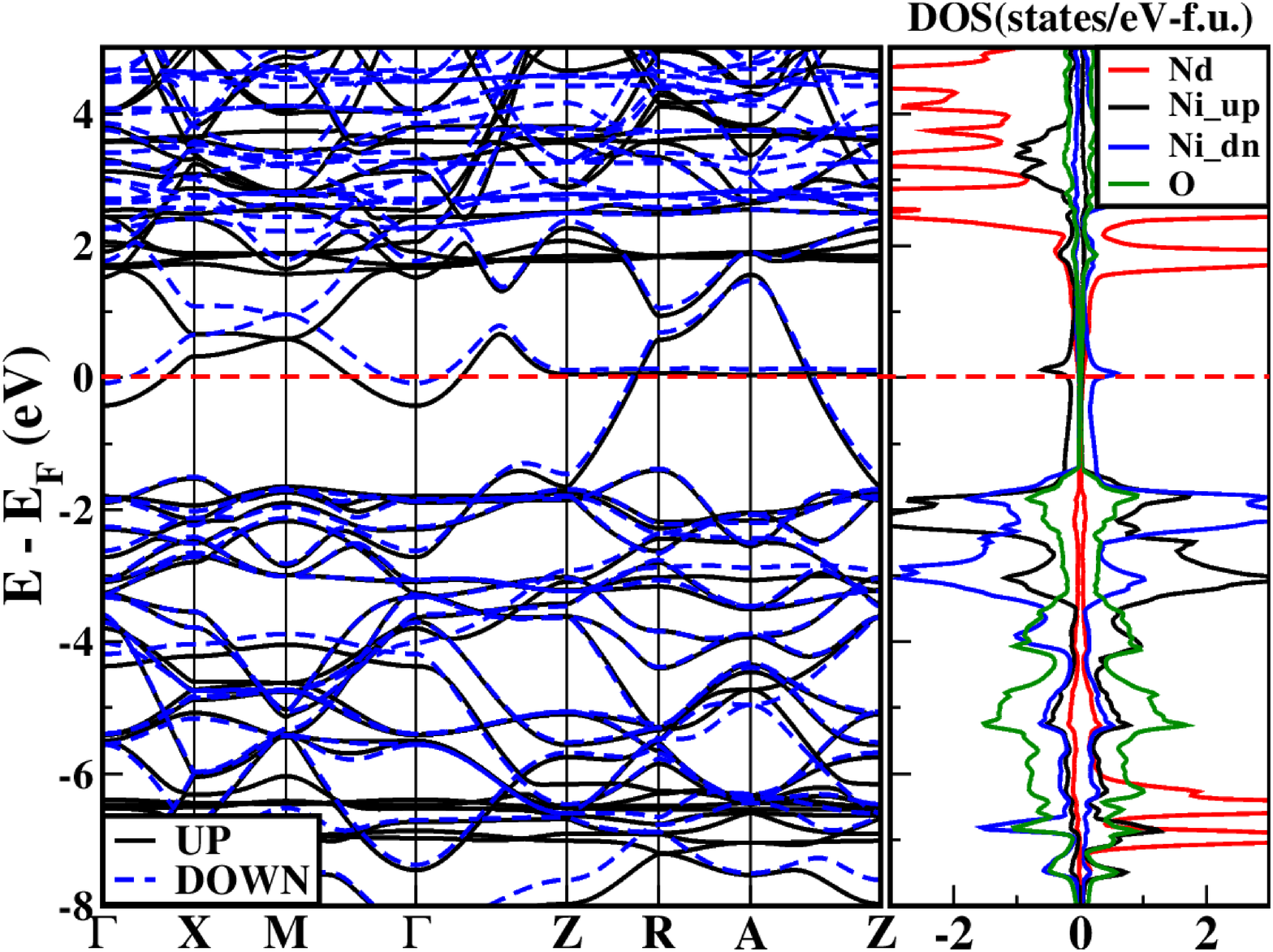}}}
\caption{GGA+U band structures and atom-projected DOSs of 
(left) AFM1 and (right) AFM2. In the AFM2 state,
a small spin-imbalance in the antiferromagnetic ordered Ni ions is 
induced by the FM ordered Nd ions
(see the region near $E_F$). Note that in both cases a flat band
along $Z-R-A-Z$ pins the Fermi level.
}
\label{afm_band}
\end{figure*}

Figure \ref{afm_band} shows the band structures along lines in the AFM zone
and atom-PDOSs of AFM1 and AFM2. 
(In the band structure plots, the {\it prime} symbol is omitted for simplicity.)
Both band structures are characterized by a near gap as a cuprate would
display, except for a band dispersing upward from $Z'-R'$ and downward from
$a'-Z'$ along the $k_z=\pi$ plane. No such dispersion is observed on the
$k_z=0$ plane $\Gamma-X'-M'-\Gamma$. The Fermi level is fixed not by 
half-filling of this band but by the lower conduction band, which is
flat along $Z'-R'-A'-Z'$. The dispersion band is of ambiguous mixed
character while the flat band has very strong Ni $d_{z^2}$ character. 

In GGA,
a magnetic Ni ion leads to a 120 meV energy gain compared to non-magnetic Ni.
This exaggerated tendency toward magnetic ordering and the magnitude of
the magnetic moment is a known deficiency of (semi)local density 
functional methods, with much of the problem attributed to the lack
of effects of spin fluctuations in the functional. 

The energies of AFM1, AFM2, and FM, obtained with the same functional,
can be compared. The energy of AFM2 (FM Nd) is
slightly lower than that of AFM1 (AFM Nd), by 7 meV for $U$=0 and
by only 1 meV (at the edge of computational precision) with $U$=5 eV
on Ni, reflecting a slight tendency toward FM ordering of the $4f$ moments.  
The energy difference between FM and AFM2 provides the difference between
AFM and FM Ni moments; AFM alignment is favored by 116 (25) meV/Ni in
GGA(+U). This energy difference contains information about Ni-Ni in-plane
exchange coupling, and would give a value of the nearest neighbor coupling
if all others were negligible. Liu {\it et al.} have derived values of
coupling for a few neighbors, concluding that the values depend strongly
on the value of $U^{Ni}$ that are used.\cite{liu2019}


\newpage

Results of AFM1 and AFM2 show similarity as expected, with differences 
in the band structures appearing as small spin-splitting of bands, most
evident in the Nd $5d$ electron-pocket band at $\Gamma$. Curiously,
this band becomes incredibly flat along the zone-top lines
$Z-R-A-Z$, pinning the Fermi level. The flatness leads to a sharp
peak at the Fermi level with the associated tendency toward instabilities
of various kinds, and might contribute to the lack of such ordering.

\begin{figure*}[htbp]
{\resizebox{12cm}{7.5cm}{\includegraphics{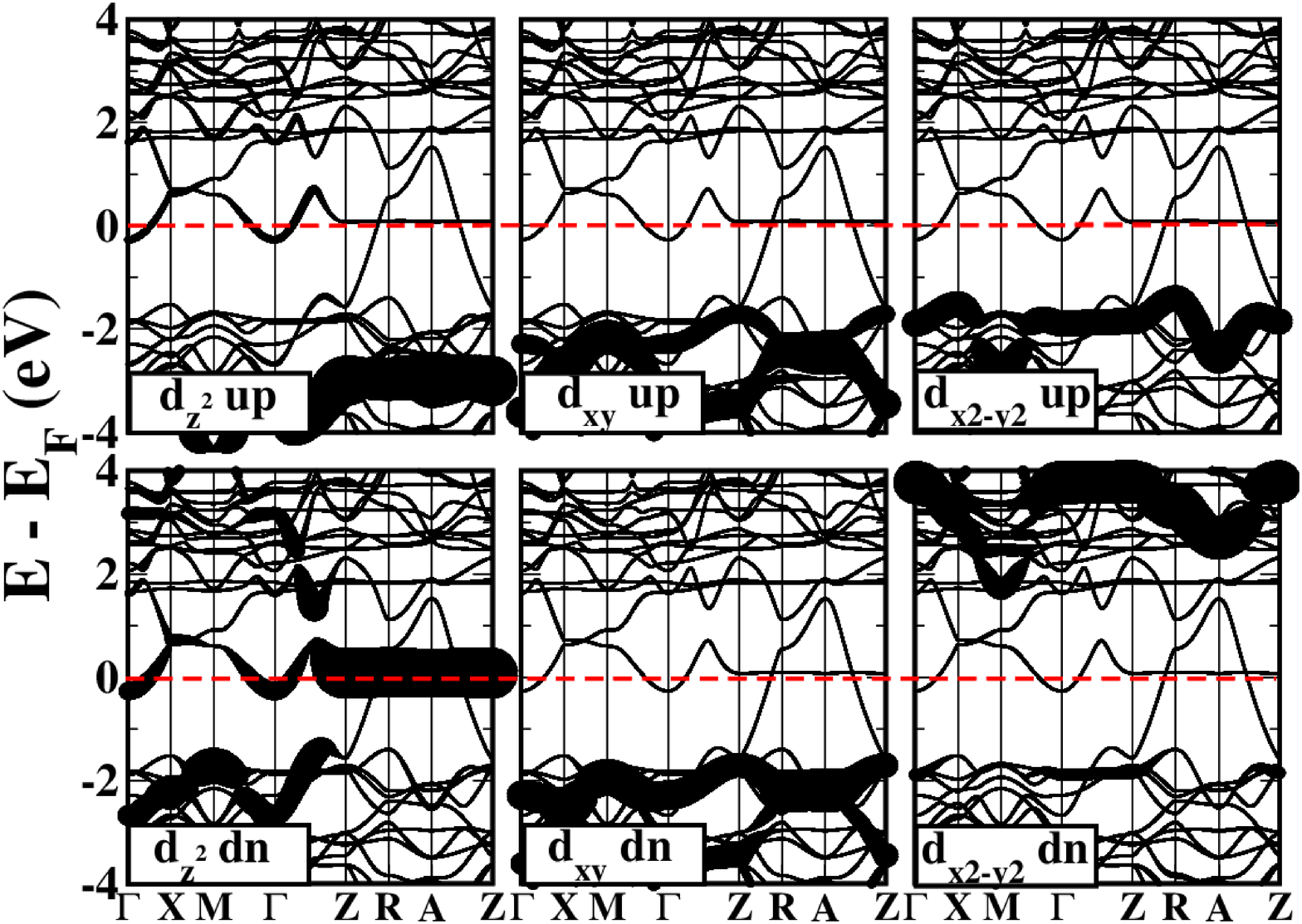}}}
\caption{AFM1 full fatband plots of Ni $3d$ orbitals in GGA+U. 
The fully occupied $d_{xz}/d_{yz}$ orbitals are not shown here.
}
\label{afmfat_3d}
\end{figure*}

\begin{figure*}[htbp]
{\resizebox{12cm}{7.5cm}{\includegraphics{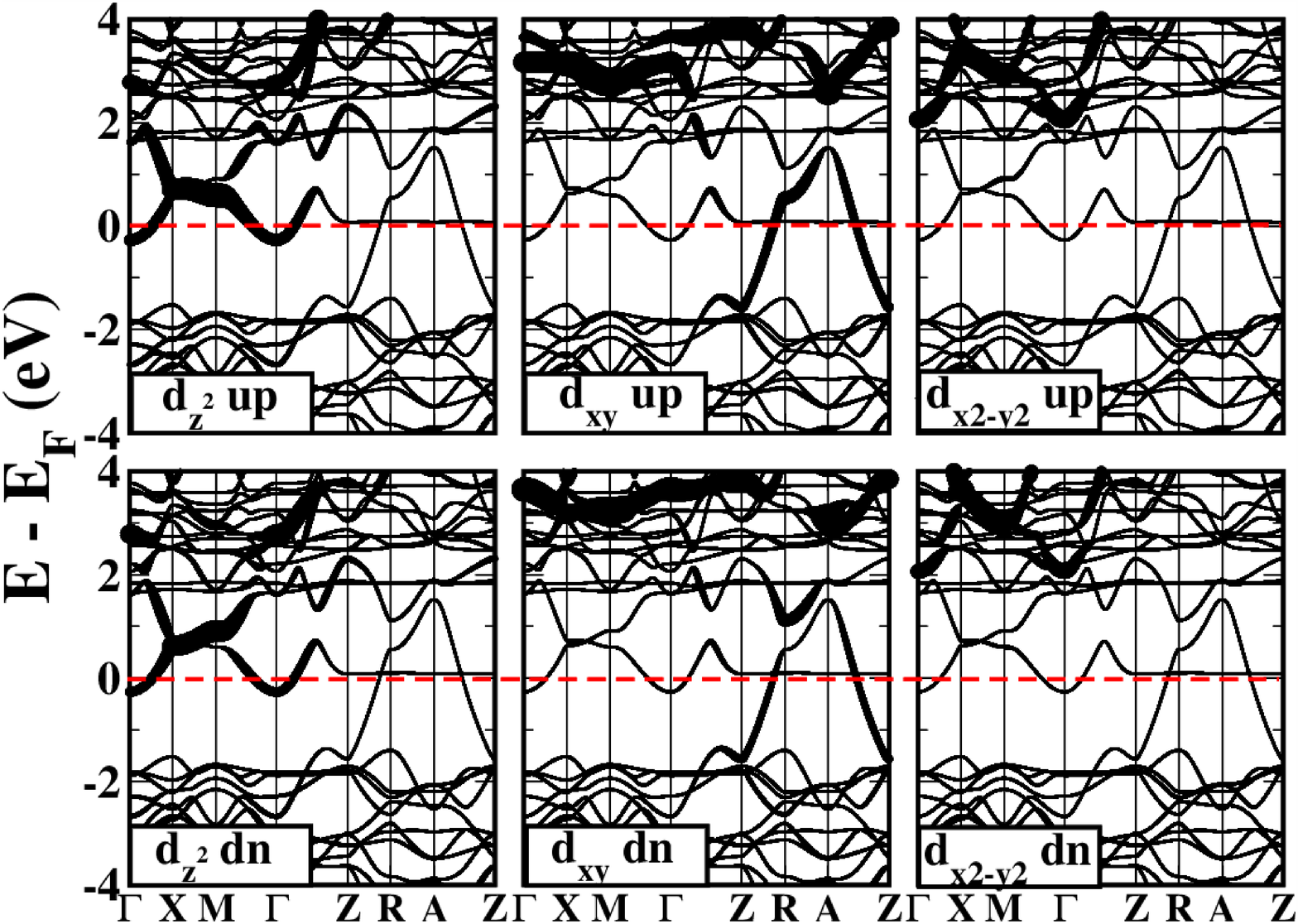}}}
\caption{AFM1 full fatband plots of Nd $5d$ orbitals in GGA+U.
The fully unfilled $d_{xz}/d_{yz}$ orbitals are not shown here.
}
\label{afmfat_5d}
\end{figure*}

\begin{figure*}[htbp]
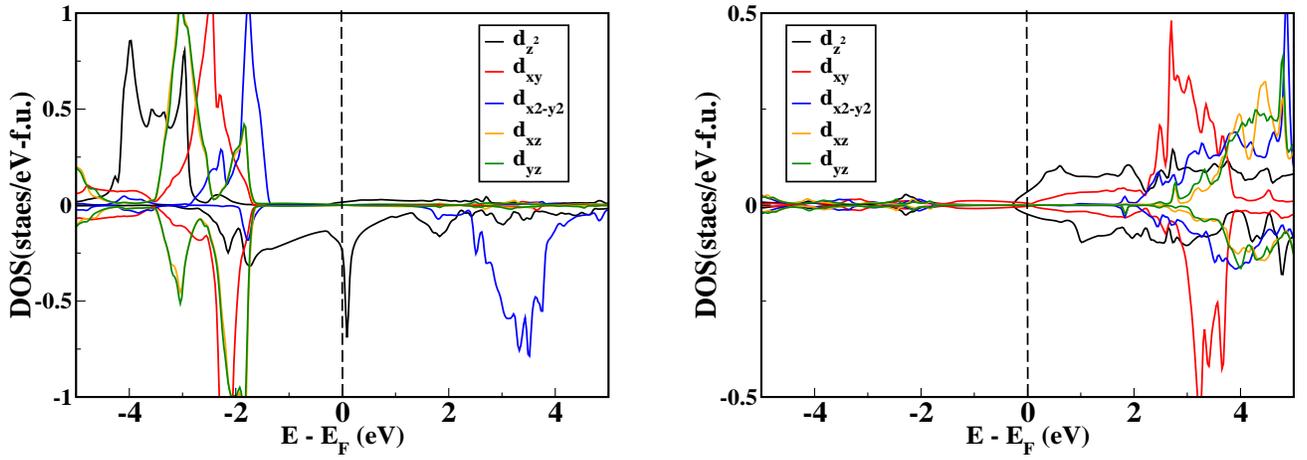

{\resizebox{8cm}{6cm}{\includegraphics{FigS7a.eps}}}
\hskip 10mm
{\resizebox{8cm}{6cm}{\includegraphics{FigS7b.eps}}}
\caption{AFM1 orbital-projected densities of states of (left) Ni $3d$ and (right) Nd $5d$ orbitals in GGA+U.
}
\label{afm_pdos}
\end{figure*}

We display the simpler band structure of AFM1,
with the fatband plots of Ni $3d$ and Nd $5d$ presented separately in 
Figs.~\ref{afmfat_3d} and \ref{afmfat_5d}, respectively. The flat band
at the Fermi level is seen to be very strongly Ni $d_{z^2}$ in
character.
The corresponding orbital-projected DOSs are given in Fig. \ref{afm_pdos},
where the Ni $d_{z^2}$ peak is very prominent at the Fermi level --
purely of a single character.

\begin{figure*}[htbp]
{\resizebox{12cm}{7.2cm}{\includegraphics{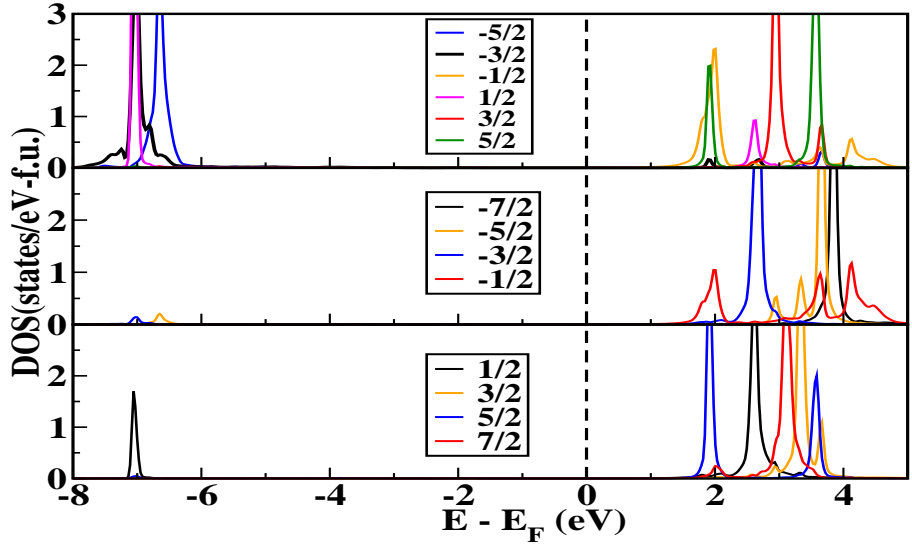}}}
\caption{AFM1 orbital-resolved DOSs of Nd $4f$ orbitals in the 
$|j,m_j>$ basis, using the GGA+U+SOC.
The occupied orbitals have mostly $|5/2,-5/2>$, $|5/2,-3/2>$, and 
$|5/2,+1/2>$ character, 
leading to the $4f$ orbital moment of --4.45 $\mu_B$. Other configurations
of similar energy might be obtained as self-consistent solutions, in
which case the orbital moment would differ.
}
\label{afm_4fdos}
\end{figure*}

\begin{figure*}[htbp]
{\resizebox{12cm}{7.8cm}{\includegraphics{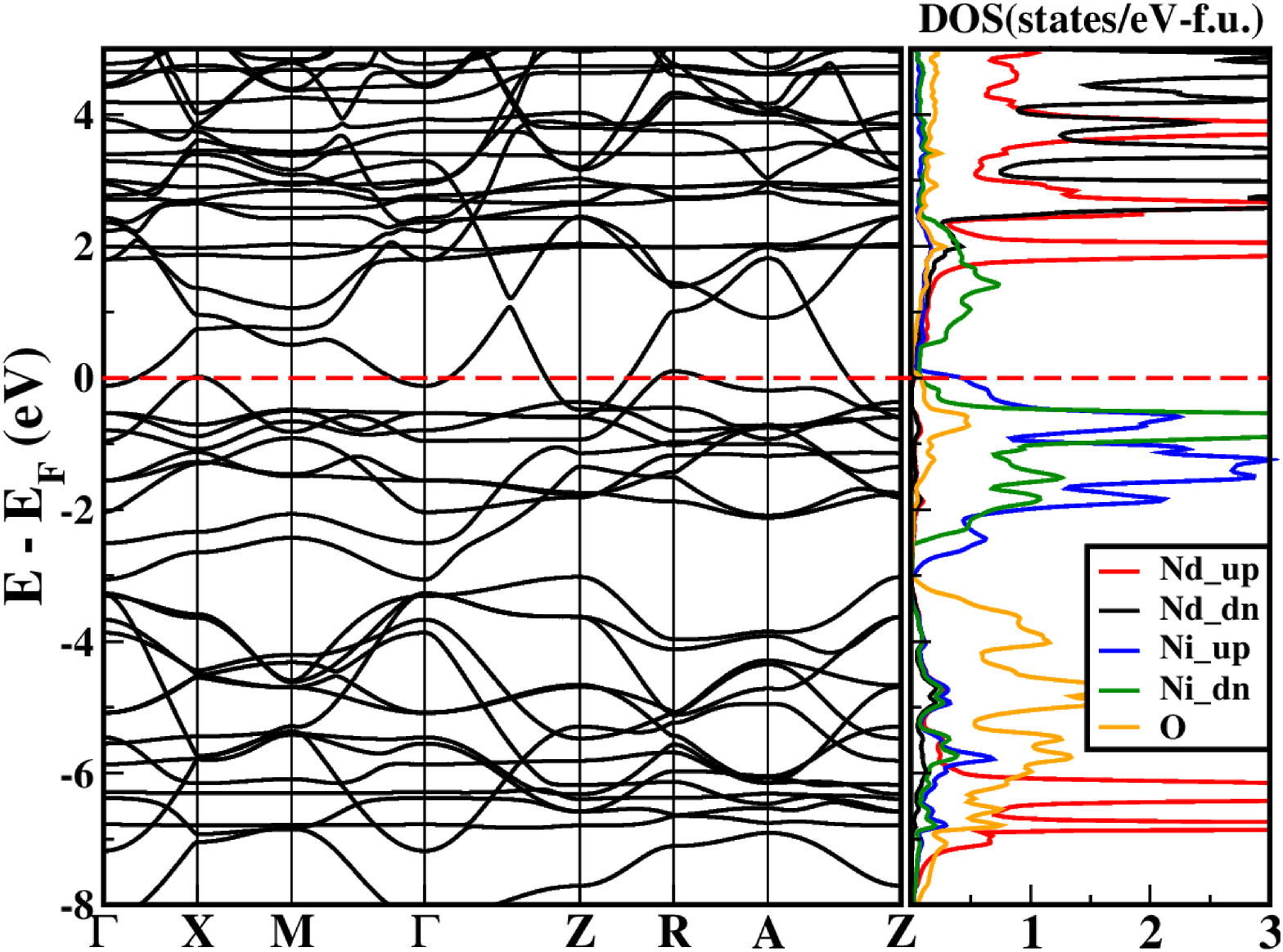}}}
\caption{AFM1 band structure and atom-projected DOSs in GGA+U, 
applying $U=8$ eV and $J=1$ eV only to the Nd $4f$ orbital.
The magnitudes of the local spin moments are Nd 3 $\mu_B$ and Ni 0.65 $\mu_B$.
Compared with that of applying $U$ to both Ni and Nd orbitals (see Fig. \ref{afm_band}),
a clear distinction appears on the $k_z=\pi$ plane (along the $Z-R-A-Z$ line), 
in addition to a few hole Fermi surface.
In this case, the Ni $3d_{z^2}$ orbital, which is dispersionless in this plane,
is fully occupied, and an electron band appears at the $Z$-point.
}
\label{afmband_nd}
\end{figure*}

We also considered the effects of SOC in the GGA+U+SOC to analyze the Nd $4f$ configuration,
leading to the Nd orbital moment of --4.45 $\mu_B$.
Figure \ref{afm_4fdos} shows the PDOS of Nd $4f$ orbitals in the $|j,m_j\rangle$ basis.
Figure \ref{afmband_nd} shows the band structure  and atom-projected
DOS, when applying $U$ and $J$ only to the Nd $4f$ orbitals
in the AFM1 alignment.

\begin{thebibliography}{10}
\bibitem{vca}The VCA was accomplished by replacing Z=60 of Nd by Z=59.8.
The small shift in $4f$ levels is unimportant.
\end{thebibliography}
\end{document}